\documentclass[sigconf]{acmart}

\usepackage{amsmath,amsfonts}
\usepackage[ruled,vlined]{algorithm2e}
\usepackage{tabularx}
\usepackage{subfig}
\usepackage{graphicx}
\usepackage{textcomp}
\usepackage{fancyhdr}
\usepackage{enumitem}
\usepackage{afterpage}
\usepackage{array}
\usepackage{bm}
\usepackage{float}
\usepackage[normalem]{ulem}
\usepackage{caption}
\usepackage{environ}
\usepackage{makecell}
\usepackage{textcomp}
\usepackage{booktabs}
\usepackage{pifont}
\newcolumntype{P}[1]{>{\centering\arraybackslash}p{#1}}
\colorlet{shadecolor}{gray!20}
\usepackage{siunitx}

\def\BibTeX{{\rm B\kern-.05em{\sc i\kern-.025em b}\kern-.08em
    T\kern-.1667em\lower.7ex\hbox{E}\kern-.125emX}}
    
\newcommand{\revision}[1]{\textcolor{black}{#1}}
\newcommand{\todo}[1]{\textcolor{black}{#1}}

\newcommand{\sol}{\textsc{MISO}}
\newcommand{\base}{\textsc{NoPart}}
\newcommand{\static}{\textsc{OptSta}}
\newcommand{\orcl}{\textsc{Oracle}}

\renewcommand\footnotetextcopyrightpermission[1]{}

\begin{document}

\title{\sol{}: Exploiting Multi-Instance GPU Capability on Multi-Tenant Systems for Machine Learning}

\author{Baolin Li}
\affiliation{%
  \institution{Northeastern University}
  \country{}}

\author{Tirthak Patel}
\affiliation{%
  \institution{Northeastern University}
  \country{}}

\author{Siddharth Samsi}
\affiliation{%
  \institution{MIT}
  \country{}}

\author{Vijay Gadepally}
\affiliation{%
  \institution{MIT}
  \country{}}

\author{Devesh Tiwari}
\affiliation{%
  \institution{Northeastern University}
  \country{}}

\renewcommand{\shortauthors}{B. Li et al.}

\begin{abstract}

GPU technology has been improving at an expedited pace in terms of size and performance, empowering HPC and AI/ML researchers to advance the scientific discovery process. However, this also leads to inefficient resource usage, as most GPU workloads, including complicated AI/ML models, are not able to utilize the GPU resources to their fullest extent -- encouraging support for GPU multi-tenancy. We propose \sol{}~\footnote{\sol{} has been accepted at the 2022 ACM Symposium on Cloud Computing (SoCC)}, a technique to exploit the Multi-Instance GPU (MIG) capability on the latest NVIDIA datacenter GPUs (e.g., A100, H100) to dynamically partition GPU resources among co-located jobs. \sol{}'s key insight is to use the lightweight, more flexible Multi-Process Service (MPS) capability to predict the best MIG partition allocation for different jobs, without incurring the overhead of implementing them during exploration. Due to its ability to utilize GPU resources more efficiently, \sol{} achieves 49\% and 16\% lower average job completion time than the unpartitioned and optimal static GPU partition schemes, respectively.

\end{abstract}
\settopmatter{printacmref=false}




\maketitle

\pagestyle{empty}

\section{Introduction}
\label{sec:intro}

\vspace{1mm}

\noindent \textbf{Background and Motivation:} Recent advancement in GPU technology has enabled HPC and AI researchers to leverage GPU computing capabilities for a wide variety of critical science missions, including training of compute-intensive neural network models~\cite{wozniak2018candle,raissi2019physics,shlomi2020graph,fung2021benchmarking}. While these advances have expedited the scientific discovery process, efficient resource utilization of the powerful GPUs remains a key bottleneck. 

With innovative progress in computing technology, GPU vendors are making individual GPUs bigger and faster -- where an individual GPU can now deliver more than 300 TeraFLOPS of performance and is on the path to becoming a supercomputer of the past by itself~\cite{a100-datasheet,chen2012looking}. This trend has served the AI/ML models well since the computing requirements of these models are increasing at a rapid pace~\cite{narayanan2021efficient,dryden2021clairvoyant,krizhevsky2012imagenet}. Unfortunately, as our experimental characterization (Sec.~\ref{sec:motiv}) and previous works~\cite{hu2021characterization,gupta2020architectural,jeon2019analysis,276938,li2022ai} have shown, even these models are not able to fully utilize the GPU computing resources, because various workloads have different resource bottlenecks and performance sensitivity to different resources. Therefore, the ``one-size-fits-all'' approach of making a single GPU more powerful is not optimal for all workloads and leads to inefficient resource utilization.\vspace{1mm}


Recognizing and motivated by these challenges, GPU vendors have recently started offering native GPU resource partitioning capabilities to enable GPU workload co-location~\cite{nvidia-mps,nvidia-mig}. These capabilities allow jobs to share the GPU resources concurrently and, thereby, reduce the cloud computing cost, reduce the long job queue wait time on HPC clusters, and potentially reduce the average job completion time (queue wait time + execution time). While promising, efficiently leveraging GPU partitioning is challenging because configuring a GPU to partition the resources optimally among co-located workloads is (1) cumbersome due to various practical partitioning constraints, (2) prohibitively time-consuming during the exploration process of finding a performance-efficient partition, and (3) incurs overhead (Sec.~\ref{sec:bkgd}). \vspace{2mm}


\textit{Therefore, the goal of this paper is to provide a novel method that automatically and quickly partitions GPU resources to achieve overall higher performance.} Solutions in this space are expected to become increasingly critical as HPC centers are beginning to deploy modern GPUs with explicit resource partitioning abilities. For example, the NVIDIA A100 GPUs, which have MIG technology support, are a part of many cloud computing offerings, industrial research computing clusters, and academic HPC centers~\cite{aws-a100,gcp-a100,top500-list,miscrosoft-blog}. But currently, we do not have the tools to leverage MIG technology to effectively utilize MIG capabilities for faster execution and higher throughput, and thereby, reducing the cost of renting GPU resources or operating HPC clusters. Our proposed solution,  \revision{\sol{}, is publicly available as an open-source package at \textit{\url{https://doi.org/10.5281/zenodo.7135988}}.}
\vspace{2mm}

\noindent \textbf{Contributions.} This paper makes the following contributions. 

\vspace{1mm}
\noindent {\textbf{I. We present experimental evidence to demonstrate the opportunities and trade-offs in GPU workload co-location capabilities provided by modern GPUs and present a novel approach to exploit this trade-off.}} In particular, our experimental characterization of the Multi-Process Service (MPS) and Multi-Instance GPU (MIG) co-location capabilities shows that they offer different levels of partition granularity and performance isolation. Our experiments further reveal that determining the optimal GPU partition for the performance of co-located jobs is non-trivial and requires extensive exploration of interference-free MIG GPU configurations -- incurring job disruption and overheads.\vspace{1mm}

\noindent {\textbf{II. \sol{} is a novel mechanism that enables efficient co-location of GPU workloads using the recent advancement in the workload co-location capabilities on modern GPUs.}} To exploit the trade-off presented by different co-location capabilities, \sol{} approaches the problem of finding optimal GPU partitions for co-located workloads with a new perspective: \textit{\sol{} proposes to use interference-prone co-location of jobs to estimate the near-optimal interference-free GPU partitions for co-located workloads.} This approach avoids the expensive exploration of different interference-free GPU partition configurations to determine the optimal partition. 

\sol{} provides a learning-based method that can accurately estimate and predict an individual job's performance on GPU interference-free partitions (MIG configurations) from quicker, more flexible but interference-prone co-location capability (MPS configurations). \sol{} then leverages this information to dynamically determine the near-optimal GPU partition for co-located job-mix. \sol{} formulates this as a practical optimization problem, and schedules co-located jobs to improve key metrics of job completion time, makespan, and system throughput.\vspace{1mm}

\noindent {\textbf{III. Our extensive real-system and simulation-based evaluation confirm that \sol{} is effective at improving the key figures of merit (e.g., job completion time, makespan, and system throughput) under different scenarios.}}  \sol{}'s experimental evaluation is driven by representative production environments~\cite{hu2021characterization} and emerging workloads such as \textit{BERT models for natural language processing, and Graph Neural Network (GNN) models for prediction of quantum chemistry molecular graphs}~\cite{gilmer2017neural}. Our real-system evaluation demonstrates and explains why \sol{} outperforms existing techniques and its effectiveness is close to the practically-infeasible Oracle technique. \sol{} outperforms the unpartitioned GPU scheme by 49\%, 15\%, and 23\% in terms of job completion time, makespan, and system throughput, respectively, and is within 10\% of the Oracle technique for all three key metrics. 

Next, we introduce the details of the state-of-the-practice GPU sharing technologies, in particular MPS and MIG, as well as the MIG capability on the latest NVIDIA A100 GPUs.

\section{Background}
\label{sec:bkgd}



In this section, we introduce and compare different GPU resource partitioning paradigms on NVIDIA GPUs. \textit{We acknowledge that we are using NVIDIA's GPU resource partitioning technology present in NVIDIA A100 GPUs as a vehicle to demonstrate the value of core ideas of \sol{}. We expect other GPU vendors to release similar capabilities in the near future as GPUs become increasingly powerful (Sec.~\ref{sec:discuss}). Along with these developments, \sol{} will continue to benefit current and future MIG-enabled cloud computing data centers and HPC centers ~\cite{aws-a100,gcp-a100,top500-list,miscrosoft-blog}.}

\subsection{GPU Resource Sharing}

Multiple applications can share one GPU using the time-sliced virtual GPU (vGPU) architecture. However, time-slicing does not address the challenge of running multiple applications that each cannot efficiently utilize a full GPU.\vspace{1.5mm}

\begin{figure}[t]
    \centering
    \includegraphics[scale=0.45]{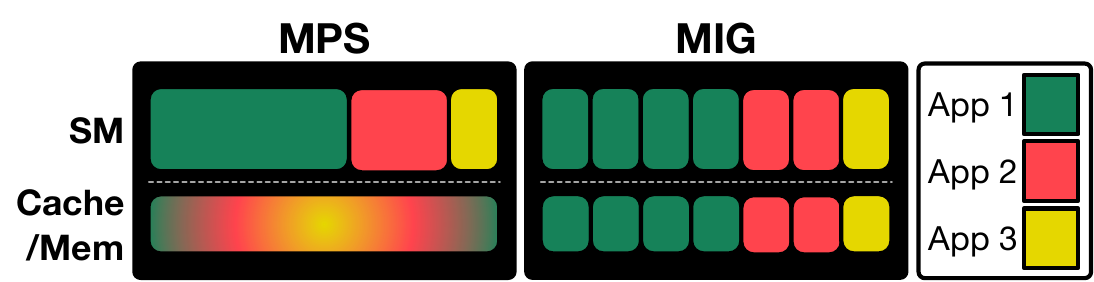}
    \vspace{1mm}
    \hrule
    \vspace{-0.4cm}
    \caption{MPS sharing mode and MIG sharing mode.}
    \vspace{-0.2cm}
    \label{fig:bkgd_1}
\end{figure}

\noindent\textbf{Multi-Process Service (MPS).} MPS is a software-based space-sharing scheme that allows applications to run on the GPU simultaneously. It partitions the GPU compute units, streaming multiprocessors (SM), into multiple partitions (represented as \% of total active threads), each partition is dedicated to a user application. MPS is the first-generation co-location support where a GPU can be configured to provide different levels of relative resource sharing among co-located workloads. \textit{The resulting co-location is not interference-free} because, as shown in  Fig.~\ref{fig:bkgd_1}, only SM resources are dedicated to each application, and the cache and memory are shared among all.\vspace{1.5mm}

\noindent\textbf{Multi-Instance GPU (MIG).} MIG is the latest hardware + software support for GPU resource sharing and partitioning supported on NVIDIA A100 Tensor Core GPUs~\cite{li2022characterizing}. MIG provides better isolation of different GPU resources among co-located workloads. Compared to MPS which only partitions the GPU SM, MIG also partitions the GPU memory, cache, and provides \revision{memory bandwidth isolation and} error isolation between concurrent applications (Fig.~\ref{fig:bkgd_1}). In other words, MIG allows the users to treat each MIG slice as a smaller A100 GPU with exclusive access, without the need to worry about performance interference with other user applications (i.e., \textit{interference-free co-location}). However, this benefit comes with some \textit{limitations}: (1) MIG only provides fixed partition sizes, the smallest partition unit on an A100 GPU with 40GB memory is \texttt{1g.5gb}, which provides 1/7 of SMs and 5GB GPU memory. MPS has a much finer granularity of SM partitions than MIG -- the user can specify the amount of SM resource using any percentage integer. (2) When a new process arrives, re-configuring MIG to make space for the new application requires stopping all applications so that the MIG slices are idle. In MPS mode, a new application can be launched if enough memory exists. 

\subsection{NVIDIA A100 Tensor Core GPUs and MIG Capability}

\begin{table}
\centering
\caption{Complete list of MIG profile on an A100 GPU~\cite{nvidia-mig} (\revision{also refer to Appendix}).}
\vspace{-0.2cm}
\scalebox{1}{
\begin{tabular}{ccccc}
\toprule
\textbf{Slice} & \textbf{Compute} & \textbf{Memory} & \textbf{Cache} & \textbf{Max Count}\\ 
\midrule
\midrule
7g.40gb & 7 GPC & 40 GB & Full & 1 \\
\hline
4g.20gb & 4 GPC & 20 GB & 4/8 & 1\\
\hline
3g.20gb & 3 GPC & 20 GB & 4/8 & 2\\
\hline
2g.10gb & 2 GPC & 10 GB & 2/8 & 3\\
\hline
1g.5gb & 1 GPC & 5 GB & 1/8 & 7\\
\bottomrule
\end{tabular}}
\label{table:bkgd}
\vspace{-0.6cm}
\end{table}

The A100 GPU's SM consists of 7 graphics processing clusters (GPC), in MIG mode, each slice (used interchangeably with MIG instance) includes at least one GPC and a corresponding amount of GPU memory. We list the full MIG slice profiles in Table~\ref{table:bkgd}. The max count means the maximum number of slices of the same type that can exist in the same GPU. The slice type notation shows the number of GPCs and the amount of GPU memory. Because the SM and memory are one-to-one mapped, we sometimes represent a slice with only the SM size (e.g., \texttt{4g}) instead of the full notation. When we mention a larger/smaller slice, it means a slice with more/less number of GPCs, respectively.

Unlike the MPS approach, arbitrary MIG partitions are not supported due to hardware restrictions. A full A100 GPU is constrained to be partitioned only into certain combinations of MIG slices. For example, both (\texttt{4g}, \texttt{2g}, \texttt{1g}) and (\texttt{2g}, \texttt{2g}, \texttt{3g}) are valid combinations. However, due to hardware limitations, some combinations cannot exist even though the resources do not exceed the A100 cap, for example, \texttt{4g.20gb} and \texttt{3g.20gb} cannot co-exist in a single A100. In total, there are 18 MIG configurations on an A100 GPU (see Appendix). For a job mix (set of jobs to co-locate), the number of configurations is large because it includes not only the configuration of the MIG hardware, but also different assignments of jobs to the created MIG slices. Each such configuration is referred to as \textit{partition configuration or MIG configuration}.

\subsection{System Throughput, Job Completion Time, and Makespan}
\label{back:metrics}


We briefly review the three widely used figures of merit relevant to quantifying the effectiveness of \sol{} and their definitions. When jobs are sharing a GPU in MIG mode, we use \textbf{system throughput}, or \textbf{STP} to measure the combined progress rate of all jobs~\cite{eyerman2008system,eyerman2014multiprogram}. This metric essentially measures how much faster the jobs are progressing towards their completions (overall progress rate), compared to when these jobs are executed one by one in exclusive GPU without co-location. Formally, for $m$ jobs $J_1$ to $J_m$, suppose job $J_i$'s execution speed on an A100 GPU without co-location is $p_i$, and its current execution speed is $q_i$ (on some MIG slice), then the system throughput is calculated as:\vspace{-2mm}

{{
\begin{align}
    \label{eq:bkgd}
    \text{System Throughput (STP)} = \sum_{i=1}^{m}  \frac{q_i}{p_i} \vspace{-1mm}
\end{align}}}

This particular definition and similar variants have been widely used in the literature for denoting system throughput~\cite{eyerman2008system,eyerman2014multiprogram}. 

\textbf{Average job completion time (JCT)} is the end-to-end service time of a job -- the sum of the time spent waiting in the queue and job execution time. Average JCT is widely used to evaluate system software in the previous works~\cite{peng2018optimus,gu2019tiresias,narayanan2020heterogeneity}, a shortened average JCT means users will experience better turnaround time and the system can support a larger user base. 

\textbf{Makespan} is the time between the start of the first job to the completion of the last job in a job trace. These three metrics will be used to extensively evaluate system performance in Sec.~\ref{sec:eval}.

\section{Motivation}
\label{sec:motiv}


In this section, first, we provide quantitative examples to demonstrate the potential benefits of partitioning GPU resources using the recently introduced Multi-Instance GPU (MIG) capability over the Multi-Process Service (MPS) method. Then, we discuss the challenges in achieving the full potential of GPU partitioning via MIG technology -- \sol{} solves these challenges. 

\begin{figure}[t]
    \centering
    \includegraphics[scale=0.52]{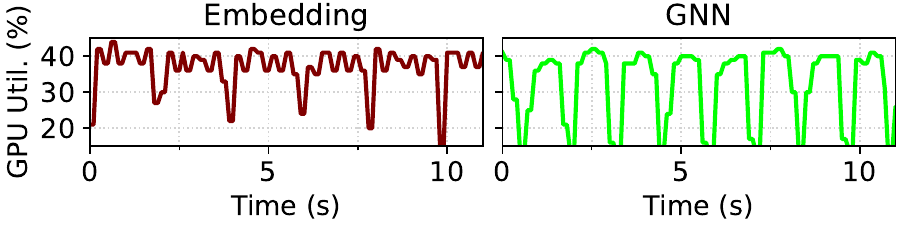}
    \hrule
    \vspace{-0.4cm}
    \caption{GPU resource utilization of example GPU applications.}
    \vspace{-0.4cm}
    \label{fig:motiv_1}
\end{figure}

\vspace{1.5mm}
\noindent\textbf{Takeaway 1. Many emerging compute-intensive workloads often cannot fully utilize compute resources in modern GPUs -- motivating the opportunity for co-location.} 

\vspace{1.5mm}
Fig.~\ref{fig:motiv_1} shows the GPU SM utilization for two representative workloads over their execution time (i.e., word embedding and graph neural network training). We note that the workloads often do not utilize the GPU resources at the maximum level. This is because modern GPUs are becoming increasingly powerful and provide higher computational power, but the workloads often have different bottlenecks (e.g., memory access latency, memory bandwidth) and hence, cannot leverage all the GPU resources to the fullest. 

\begin{figure}[t]
    \centering
    \includegraphics[scale=0.51]{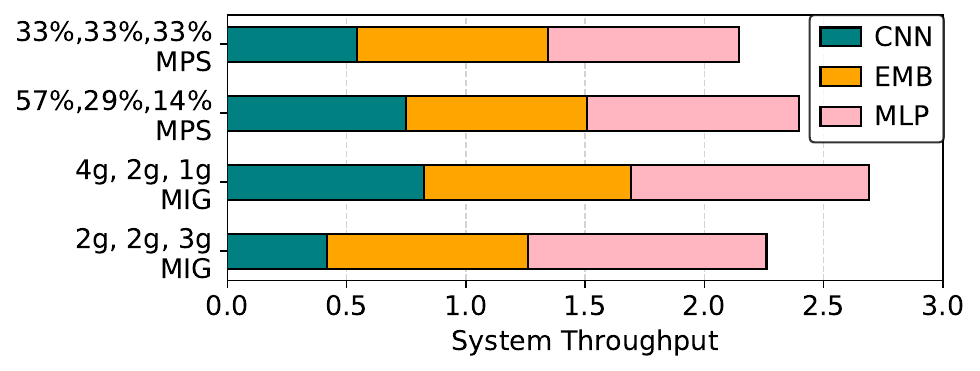}
    \hrule
    \vspace{-0.4cm}
    \caption{The system throughput of a workload mix with MPS sharing \revision{(top two bars)} and MIG sharing \revision{(bottom two bars)} on A100s.}
    \vspace{-0.4cm}
    \label{fig:motiv_2}
\end{figure}

\vspace{1.0mm}
\noindent\textbf{Takeaway 2. MIG's capability for workload co-location provides further opportunity for performance improvement beyond the MPS' method of co-location.} 

\vspace{1.0mm}
Fig.~\ref{fig:motiv_2} shows the overall performance observed when three jobs (CNN, embedding, and multi-layer perceptron) are co-located as a job mix on an A100 NVIDIA GPU. The overall performance is indicated as system throughput (Eq.~\ref{eq:bkgd}). First, we note that MPS-enabled co-location (first bar) allows multiple jobs to run together, and hence, achieve higher throughput than what would have been possible if co-located jobs were sequential (STP = 1). Second, we note that a MIG configuration (third bar) can provide higher system throughput than the MPS co-location. However, our MPS co-location (33\%, 33\%, 33\%) shared the resources equally among co-located jobs, but the MIG partition (\texttt{4g}, \texttt{2g}, \texttt{1g}) divides the GPU computing resources in the ratio of 4:2:1. For a fairer comparison, we configure the MPS scheme to share the resources in the same proportion (second bar) and noticed that the MIG partition still yields higher performance. \revision{This is because the two workloads CNN and EMB both have seen improved performance even though the SM resources are the same as MPS, underlining the MIG's benefit of performance isolation and resource exclusivity among co-located jobs (illustrated in Fig.~\ref{fig:bkgd_1}).} \revision{We note that not all MIG configurations can outperform MPS configurations. A poorly-chosen MIG’s system throughput (e.g., a workload needing the smallest memory capacity but assigned the largest MIG slice) will underperform MPS. For example, the (57\%, 29\%, 14\%) MPS partition outperforms the (\texttt{2g}, \texttt{2g}, \texttt{3g}) MIG partition. However, MIG provides control knobs for partitioning different architectural resources, while MPS only controls the SMs and cannot control interference among co-located workloads for other resources (memory, bandwidth, etc.). Therefore, MIG is expected to outperform MPS in most cases. Our motivational example serves the simple purpose of demonstrating that better isolation achieved via MIG configurations can lead to higher performance.}

\begin{figure}[t]
    \centering
    \includegraphics[scale=0.46]{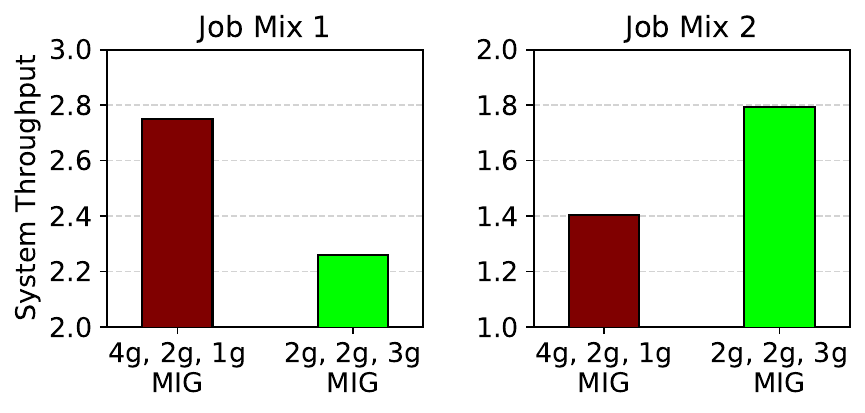}
    \hrule
    \vspace{-0.4cm}
    \caption{Sharing the GPU with different MIG partition patterns results in different performances (left). When the job mix changes, the optimal MIG partition may also change (right). Job mix 1 consists of (CNN, EMB, and MLP), while job mix 2 consists of (MLP, DeepSpeech, and GNN).}
    \vspace{-0.4cm}
    \label{fig:motiv_3}
\end{figure}

\vspace{1.5mm}
\noindent\textbf{Takeaway 3. Optimal MIG partition configuration among workloads changes across different job mixes, but exploring for the optimal partition incurs prohibitive overhead -- this is due to frequent GPU reconfiguration, high number of GPU resets, and I/O overhead due to repeated workload checkpoint-and-restart.} \vspace{1.5mm}

Fig.~\ref{fig:motiv_3} shows the system throughput for two different job mixes running on two different MIG partition configurations. As one would expect, different partitions result in different performances for the same job mix. More interestingly, the performance ordering of the two MIG configurations is inverted for different job mixes. Therefore, when different mixes of workloads are co-located, the optimal resource configuration is likely to be different. It is critical to find the optimal GPU partition for a given workload mix. 

Unfortunately, finding the optimal GPU partition for a given workload mix is challenging because it requires experimentally evaluating the performance corresponding to different MIG partition configurations, which is time- and cost-prohibitive. 

First, the number of possible MIG partition configurations is many and each configuration needs to be in effect for a certain duration for estimating its corresponding performance with high confidence. Second, each MIG configuration performance evaluation requires resetting the GPU, hence, disrupting the progress of all co-located jobs (it takes about 4 seconds for each GPU MIG reconfiguration). All jobs need to be restored back to their execution state when a new MIG configuration is put in effect. This requirement generates additional time and I/O overhead. The corresponding checkpoint overhead and the application restart time after MIG reconfiguration can be from seconds to minutes in practice. \textit{In contrast, exploring different resource sharing levels of a job in MPS mode does not disrupt the execution of other jobs, all jobs in the GPU can execute concurrently in any MPS configuration.}

\begin{figure}[t]
    \centering
    \includegraphics[scale=0.46]{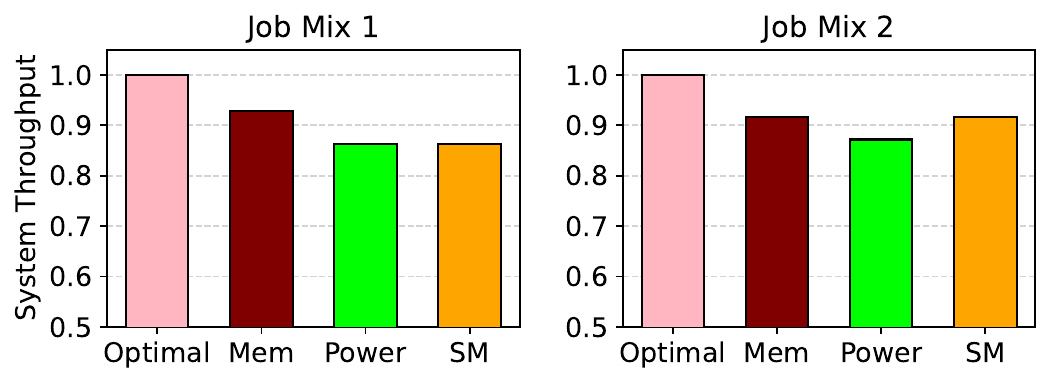}
    \hrule
    \vspace{-0.4cm}
    \caption{\revision{Applying heuristic-based approaches to perform MIG partition (using job's memory consumption, GPU power consumption, and SM utilization) does not yield the optimal MIG partitioning.}}
    \vspace{-0.5cm}
    \label{fig:motiv_4}
\end{figure}

\revision{One can apply heuristic-based methods to avoid this evaluation process, but our experimental results show that such heuristic-based methods do not always guarantee to find an optimal partition. We design the heuristic to partition the GPU according to the job memory, GPU power consumption or SM utilization of each job when running exclusively on A100 GPUs. For each method (memory, power, or SM), we use the MIG partition whose number of GPCs has the highest Cosine similarity to the collected characteristic of the job mix. For example, if jobs in a job mix have memory sizes of 4000MB, 2500MB, and 1000MB, then we assign the partition (\texttt{4g}, \texttt{2g}, \texttt{1g}) to it because [4,2,1] has the highest Cosine similarity with [4000,2500,1000] than other partitions. Fig.~\ref{fig:motiv_4} shows two examples where using the heuristic-based method to partition the GPU yields 8\% to 14\% lower system throughput than the optimal partition.} \vspace{2mm}

\textit{\textbf{\sol{} takes a novel approach that combines the best of both worlds of MPS and MIG}} -- \sol{} estimates the overall performance of different MIG configurations via quickly configurable MPS resource sharing levels instead of experimentally evaluating the MIG configurations exhaustively. This allows \sol{} to avoid the overhead of exploring different interference-free MIG resource partitions. Instead, \sol{} leverages the quick and more flexible, but interference-prone MPS GPU sharing mechanism to project the performance in the interference-free GPU partition situation (MIG). Ultimately, \sol{} uses this knowledge to determine the near-optimal MIG GPU resource partition to yield higher throughput and lower job completion time (more performance-efficient colocation reduces the job queue wait time).

\section{\sol{}: the \underline{MI}g \underline{SO}lution}
\label{sec:desi}

\begin{figure}[t]
    \centering
    \includegraphics[scale=0.39]{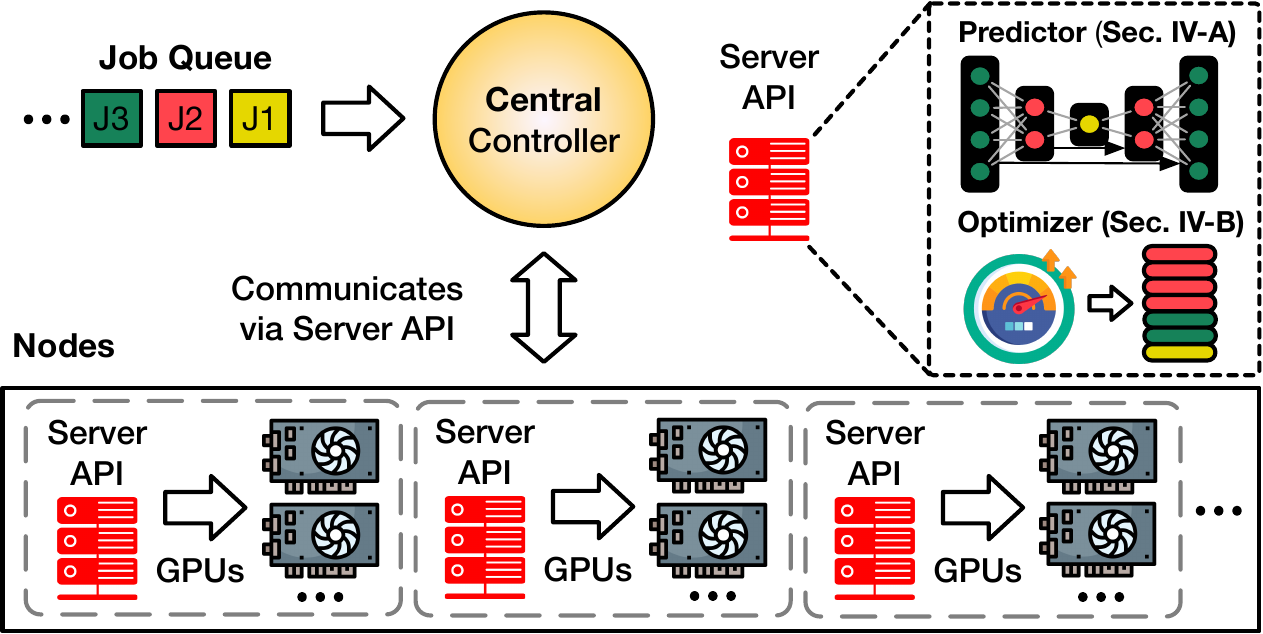}
    \vspace{1mm}
    \hrule
    \vspace{-0.4cm}
    \caption{\sol{} design overview.}
    \vspace{-0.4cm}
    \label{fig:desi_1}
\end{figure}

Fig.~\ref{fig:desi_1} shows an overview of \sol{}. \sol{} uses a central controller that monitors submissions from a job queue and communicates with server APIs distributed across nodes for status updates and scheduling decisions. Each server API corresponds to one MIG-enabled GPU. \sol{} uses a performance predictor (Sec.~\ref{sec:desi1}) to estimate every job's performance in a given job mix (set of co-located jobs) on different MIG slices using a learning-based predictor. It does so without running the job in the expensive, isolation-free MIG mode, instead, the co-located jobs are run only in the flexible, no-overhead interference-prone MPS mode only. Then, \sol{} uses these performance estimations to determine a MIG partition to maximize the overall performance, formulated as an optimization problem (Sec.~\ref{sec:desi2}).

\subsection{{\sol{} Performance Predictor}}
\label{sec:desi1}

\begin{figure}[t]
    \centering
    \includegraphics[scale=0.41]{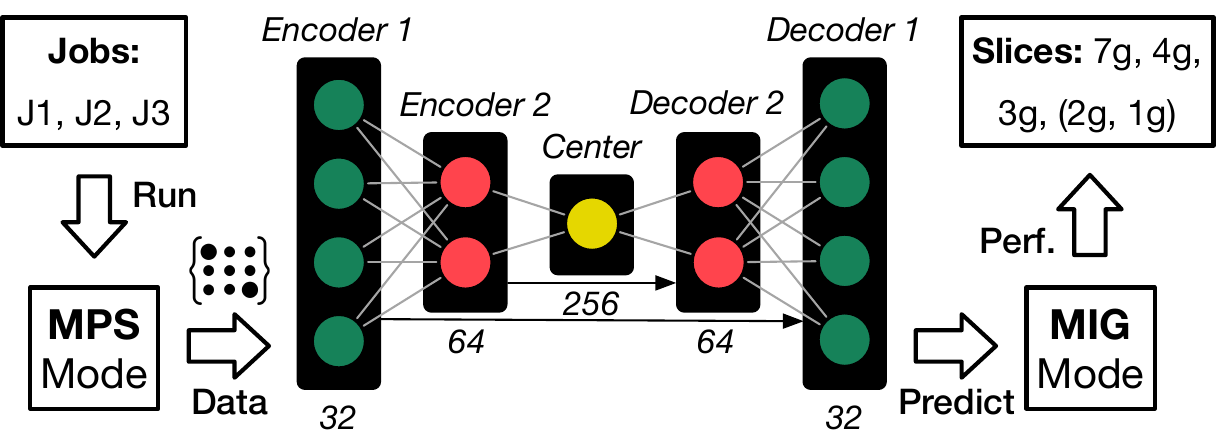}
    \vspace{-3mm}
    \hrule
    \vspace{-0.4cm}
    \caption{\sol{} predictor to translate MPS performance to MIG.}
    \vspace{-0.4cm}
    \label{fig:desi_2}
\end{figure}

\vspace{2mm}\textbf{MPS-to-MIG Performance Estimation.} The \sol{} predictor estimates a job's execution speed on different MIG slices (GPU partitions) relative to the maximum speed possible (i.e., when the job is run on the A100 full slice: \texttt{7g.40gb}). The key constraint is that \sol{} should not exhaustively run a given job on all possible GPU partitions (MIG slices) to generate the performance estimations for different MIG slices, because doing so would require frequently switching each job in the job-mix in and out of the GPU, incurring significant overhead and job idle time. To solve this challenge, \sol{} adopts a learning-based approach to build a model for predicting a job's performance on all MIG slice types. The key idea is illustrated in Fig.~\ref{fig:desi_2}. 

At first, it might appear natural to train the learning-based model with different types of jobs in all possible MIG modes to make performance estimates on different MIG slices. However, recall that during the model-inference stage, we can not provide all the job performances on different MIG slices since that would require running each co-located job separately in interference-free MIG mode and incur GPU reset and checkpoint/restart overheads. Collecting these features is detrimental because jobs have to take turns to be profiled, while the other co-located jobs have to be stopped to make space for them. For example, assume there are five co-located jobs J1-J5 on an A100 GPU, to profile J1 on \texttt{7g}, J2-J5 have to be paused. Similarly, for \texttt{4g}/\texttt{3g} profiling, jobs have to take turns to be profiled, and the accumulated waiting time adds up. 

Instead, \sol{} runs co-located jobs together in the MPS mode; then, it generates the model input features that are required to be used during the \sol{}'s model inference stage; then, the MPS-to-MIG performance estimation for each job in the job mix is combined to determine an effective partition configuration for the given job mix using a scheduler optimizer (Sec.~\ref{sec:desi2}). To confirm this experimentally, we measured the total profiling time for the number of co-located jobs using MIG-based profiling, which incurs up to 8$\times$ more overhead than \sol{}'s MPS mode profiling (in orders of minutes), to achieve similar scheduling quality as obtained by \sol{}. Also, as expected, MIG-based profiling gets worse as the number of jobs increases. In contrast, \sol{} retains near-constant cost due to concurrent execution of co-located jobs in MPS mode (shared contention-prone execution, but no GPU resets, no multiple rounds of evaluation, and fixed checkpoint-restart). In summary, \sol{} is more attractive than the MIG-based profiling because the MIG-based profiling requires frequent GPU resets and requires multiple rounds of evaluation since not all jobs can be profiled concurrently (and hence, multiple rounds of checkpoint-restart overhead and wait time). 



\textit{Next, we discuss the key design trade-offs and lessons learned in designing the ML-based \sol{} performance predictor. \sol{}'s model should be able to estimate performance, not only the relative ranking, on all MIG slice types (interference-free execution with different resource configurations) with only interference-prone runs in the MPS mode (no performance isolation).} We observed that translating to MIG performance from MPS runs requires us to be able to extract per-job interference-free high-level features from the interference-prone MPS profile. The interference-free high-level representation is needed because MIG provides hardware-level isolation between jobs. Using this as a motivation, \sol{} 's design employs an autoencoder-based neural network because the center of the autoencoder represents the key abstract features. For example, we learned via experiments that collaborative filtering, widely used by previous resource schedulers~\cite{delimitrou2014quasar,narayanan2020heterogeneity} is not suitable because they only produce the relative ranking, and other ML techniques such as linear regression, regression trees, and multi-layer perceptrons were not effective because they were unable to converge to an accurate state with limited input features.\newline

\noindent\textbf{Predictor Design.}  We construct a variant of the U-Net~\cite{ronneberger2015u} convolutional autoencoder model. It is a lightweight model with fewer encoder/decoder blocks and fewer convolutional filters compared to typical models used in applications~\cite{veillette2020sevir}. As shown in Fig.~\ref{fig:desi_2}, the input is passed through two encoder blocks with 32 and 64 convolutional filters into its center with 256 filters, then through two decoder blocks into the translated MIG speedups. The convolutional filter size is 2$\times$2 and the strides are (2,2) in horizontal and vertical directions.\vspace{2mm}

\begin{figure}[t]
    \centering
    \includegraphics[scale=0.4]{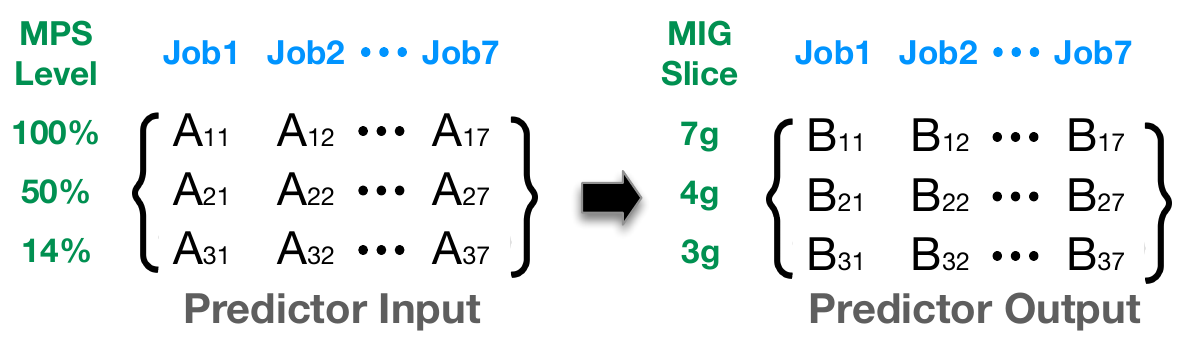}
    \hrule
    \vspace{-0.4cm}
    \caption{\revision{Input and output of the ML-based predictor.}}
    \vspace{-0.4cm}
    \label{fig:desi_inout}
\end{figure}

\textit{Input and output.} \revision{The inputs and outputs of the U-Net model are summarized in Fig.~\ref{fig:desi_inout}.} The input to the model is a 3$\times$7 matrix collected from MPS, corresponding to 3 MPS levels and 7 jobs running concurrently. The output from the model is also a 3$\times$7 matrix, each column maps to a job, and the 3 rows represent performance on the \texttt{7g}, \texttt{4g}, and \texttt{3g} MIG slices. For both the input and the output, each job (column) represents its execution speed at different MPS levels/MIG slices, normalized by the maximum speed in that column; all elements are within $(0,1]$.

We set MPS active SM for all jobs at three different percentage levels: 100, 50, and 14. The intuition is to vary the amount of SM resources shared by jobs during MPS: at 100, all jobs share access to the full GPU; at 14, all 7 jobs have their own exclusive SM block partitioned by MPS; at 50, it is a middle ground between fully shared GPU SM and exclusive SM for every job. We set 7 as the number of columns (jobs) because the A100 GPU allows a maximum of 7 jobs in MIG mode. At each knob level, we profile the job execution speed for 10 seconds. An example of this speed is the number of mini-batches per second in the training of AI/ML applications. 

Since the prediction model always takes 7 jobs (columns) as input to run in MPS, when there are less than 7 jobs, we pad the job mix with lightweight dummy workloads that we create until there are 7 total workloads. We use dummy workload padding instead of padding the input matrix with new columns of 0's because we find that large areas of zero padding greatly increase the training loss.

\vspace{2mm}
\textit{Memory considerations.} Notice that the output only contains speedup information on \texttt{7g}, \texttt{4g}, and \texttt{3g} slices. This is because some jobs cannot fit in the memory of \texttt{2g} and \texttt{1g}, while all MIG-compatible jobs will fit into \texttt{4g} and \texttt{3g} slices as both of them have the largest memory (20GB) of partitioned slices (\texttt{7g} is unpartitioned). For jobs that can fit into the \texttt{2g} or \texttt{1g} slices, we find that as long as we have the output on \texttt{7g}, \texttt{4g}, and \texttt{3g} slices, the \texttt{2g} and \texttt{1g} output can be accurately predicted by a linear regression model from the other three slice types with an $R^2$ score of 0.96. \revision{Here $R^2\in [0,1]$ is coefficient of determination, where $R^2=1$ means the regression model can explain all the variations in the data perfectly.}\vspace{2mm}

\noindent\textbf{Model training. } To train the U-Net model, first, we need to collect training data for random job mixes running on both MPS and MIG modes. The data is collected by running randomly generated workload mixes (details in Sec.~\ref{sec:metho}), whose job count ranges from 1 to 7. We generate 400 job mixes for each job count number, so in total, we have created 2800 job mixes for training. Each job mix is represented as a 3$\times$7 MPS matrix input (with dummy filling) and a 3$\times$7 MIG matrix as the target. We also perform data augmentation using the fact that the same set of jobs can be represented in different orders in the input/output matrix, but their MPS/MIG speedups will not change. Therefore, we create four extra different column permutations for each job mix -- making the total data count 14,000. From the 14,000 data points, we randomly select 75\% as training data and the rest as the validation set. 

We train the model with mean absolute error (MAE) loss and Adam optimizer~\cite{kingma2014adam}. These hyperparameters along with others such as learning rate and activation function are selected using the ASHA hyperparameter tuning algorithm~\cite{li2020system} on Ray Tune~\cite{liaw2018tune}. The validation loss converges quickly: we train the model for 50 epochs, and the validation loss (MAE) is 0.017, which is 1.7\% over the MIG speedup target range. The training is also speedy: each epoch takes 3 seconds on one A100 GPU.

\subsection{{\sol{} Scheduling Optimizer}}
\label{sec:desi2}


Now, we introduce the second component of \sol{} -- which determines the GPU partitions (MIG slices) for a given job mix using the performance data collected purely in the MPS and the \sol{} performance predictor. When a new job is scheduled on a GPU, the GPU goes into MPS mode to profile the current job mix, and \sol{} solves an optimization problem to generate the new optimal partition configuration. First, we demonstrate how \sol{} formulates and solves the partition problem for a given job mix on a single GPU, and then, discuss the trade-offs in scaling it to the cluster setting.\vspace{2mm}

\noindent\textbf{GPU resource partitioning optimization.} The goal is to use the MPS performance data collected for a short duration and performance predictor to determine the MIG partition for each job in a job mix on a single GPU. Suppose there are $m$ jobs $J_1$ to $J_m$ ($m \leq 7$ because each A100 can be partitioned into at most 7 slices), the partition configuration as the optimization variable can be represented as $\bm{\vec{x}} = [x_1, x_2, ..., x_m]$ which has the same number of elements as number of jobs, $x_i$ represents the MIG slice job $J_i$ runs on. Since MIG partitions GPUs into pre-defined chunk sizes, $x_i \in \{1,2,3,4,7\}$, where each number corresponds to a unique MIG slice type (e.g., 1 means \texttt{1g}, 7 means \texttt{7g}). We do not make any assumptions about job execution time in the optimizer, in fact, when a job will run to completion is often difficult to predict~\cite{gu2019tiresias,hu2021characterization}. Hence, we judge the merit of a particular configuration $x_i$ by the sum of each job's execution speed normalized to their maximum speed when running on an exclusive GPU. Our goal is to maximize the total system throughput for $\bm{\vec{x}}$. Without the job speedup information on each slice type, this is a black-box optimization problem: we cannot know the performance of each configuration $\bm{\vec{x}}$ unless we experimentally partition the GPU and assign the jobs on their corresponding slice to evaluate the overall performance. This is infeasible because every time we re-partition the MIG, checkpointing overhead occurs for the jobs. \textit{This is why the performance estimator performs a key role for \sol{} -- we do not need to keep reconfiguring and re-partitioning the jobs and the GPU during our optimization process.}

\begin{figure}[t]
    \centering
    \includegraphics[scale=0.46]{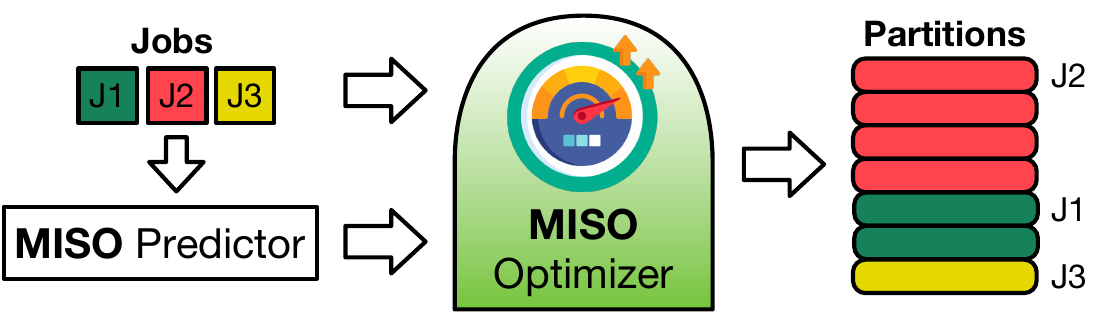}
    \hrule
    \vspace{-0.4cm}
    \caption{\sol{} optimizer to generate optimal MIG partition.}
    \vspace{-0.4cm}
    \label{fig:desi_3}
\end{figure}

Fig.~\ref{fig:desi_3} shows an overview of the \sol{} optimizer that runs for each GPU. This optimizer is run to re-partition each MIG-enabled GPU whenever a new job starts on the GPU (during MPS), or whenever a job has finished execution to ensure that the GPU has no unused MIG slice at all times. With job speedup information on each slice type as input, the optimizer can immediately find the optimal partition without interrupting job execution during the process. The profiled information for each job $i$ is represented as a function $f_i: x_i \rightarrow k_i$, where $k_i \in (0,1]$ is the job execution speed on MIG slice corresponding to $x_i$, normalized by the maximum speed on slice \texttt{7g.40gb}. For example, if $J_1$ runs 50\% slower on \texttt{3g.20gb} compared to the full GPU, $f_1(3) = 0.5$. The problem is as follows:\vspace{-4mm}

{
\begin{align}
    \label{eq:desi_obj}
    &\max_{\bm{\vec{x}}}
    \begin{aligned}[t]
      & \sum_{i=1}^{m}  f_i(x_i)
    \end{aligned} \\
    & \label{eq:desi_c0}\text{s.t. } \;\bm{\vec{x}} \in P_{mig}\\
    & \label{eq:desi_c1} \;\;\;\;\;\;\lVert\bm{\vec{x}}\rVert_0 = m    
\end{align}}

Here $P_{mig}$ represents all the available partition configurations on an A100 GPU (from MIG documentation~\cite{nvidia-mig}). For example, $\bm{\vec{x}} = [4,2,1]$ is a feasible MIG partition, so is $[4,1,2]$ because the physical partition is the same, the difference is that $J_2$ and $J_3$ are mapped to different slices. Thus, we use Eq.~\ref{eq:desi_c0} to guarantee that the partition configuration is feasible. Eq.~\ref{eq:desi_c1} means that the partition must have the same number of slices as the number of jobs -- no slice bubbles or unscheduled jobs.

Algorithm~\ref{algo:miso} shows the pseudo-code that runs at each GPU upon job start and job completion. Because of its simplicity and lightweight, we do not observe any negative impact on the running jobs. The maximum optimizer runtime during our experiments is 0.5ms, negligible compared to jobs that run for orders of magnitude longer.\vspace{2mm}

\noindent\textbf{\sol{} for cluster setting.} Optimizing job assignments on a MIG-enabled GPU cluster introduces a new dimension of complexity. Suppose there are $n$ GPUs in the system, when scheduling jobs onto the $n$ GPUs with MIG enabled, one needs to consider how to partition each GPU. Each A100 GPU can be partitioned in 18 different ways~\cite{nvidia-mig}, thus the MIG configuration space is $O(18^n)$, which is exponential. 

Therefore, globally configuring the MIG partitions across the whole cluster is a non-polynomial (NP) problem. Instead of tackling this NP problem, which could result in response time violation from the optimizer on large-scale systems, \sol{} simplifies it by locally solving a polynomial problem at every GPU. The reason it becomes polynomially solvable at each individual GPU is that the number of MIG configurations is capped at 18, and the number of jobs is capped at 7. One may hypothesize that this approach would miss out on the opportunity to migrate jobs among GPUs. However, based on our empirical experience, the performance gain from moving jobs between GPUs globally is not necessarily beneficial compared to the overhead. One overhead is from solving an NP-hard problem. 

The other major source \revision{of} overhead is from extra checkpointing: when moving a job $J_1$ from GPU A to GPU B, GPU A needs to be re-partitioned so the jobs co-located with $J_1$ can access its resources; GPU B also needs to be re-partitioned to make space for $J_1$. Thus, all other jobs in GPU A and B will be checkpoint-restarted, causing systemic overhead. The performance gain from a better global configuration diminishes with the interruption of more jobs. In fact, our evaluation shows that even one-time checkpointing overhead can be significant enough. 

\begin{algorithm}[t]
\SetAlgoLined
$best\_obj \leftarrow 0$ \tcp{Maximum objective so far}
$best\_config \leftarrow None$ \tcp{Best partition so far}
$P_{valid} \leftarrow $ list of $P_{mig}$ partitions whose length equals $m$ \\
\ForEach{$\bm{\vec{x}}$ \text{\textbf{in}} $P_{valid}$}{
    $obj\_func \leftarrow \sum_{i=1}^{m}  f_i(x_i)$ \\
    \If{$obj\_func > best\_obj$}{
        $best\_obj \leftarrow obj\_func$ \\
        $best\_config \leftarrow \bm{\vec{x}}$ \\
    }
}
\Return $best\_config$\\
\caption{MISO's partition optimizer.}
\label{algo:miso}
\end{algorithm}

\subsection{Miscellaneous Design Considerations}
\label{sec:desi3}


\vspace{2mm}\noindent\textbf{Initial job placement and dynamic adaptivity.} \sol{} monitors a first-come-first-serve (FCFS) queue and minimizes checkpointing overhead. It schedules a new job on the GPU that is hosting the least number of jobs. This policy aims to cause the least amount of disruption to all the jobs that are currently running in the cluster. When a new job is scheduled, the host GPU needs to go into the MPS mode for profiling, thus all jobs currently sharing the GPU in MIG mode will need to be checkpoint-restarted to run on MPS. Upon the profiling completion, the process repeats as the GPU switches from MPS back to MIG. Since the new job's execution characteristics are still unknown upon arrival, \sol{} attempts to minimize the negative impact on already running jobs. 

We note that starting new jobs on the least crowded GPU helps with load balancing -- all the GPUs in the cluster will host a similar number of jobs. It prevents the pathological case where multiple jobs are contesting for the resource of certain GPUs while other GPUs are underutilized. If \sol{} detects a significant change in execution speed for a running job (e.g., phase change), it will treat it as a new job and starts the MPS process for better repartition. \sol{} maintains configurable thresholds and historical data to ensure that re-invocations balance the trade-off between invocation cost and corresponding performance benefit from repartitioning.\vspace{1.mm}

\vspace{2mm}\noindent\textbf{Job out-of-memory.} Different MIG slices provide different GPU memory sizes; some jobs may face out-of-memory errors when running on smaller slices. Users may specify the minimum GPU memory needed for each job. During the MPS stage, \sol{} also monitors the GPU memory usage for individual jobs using the \texttt{nvidia-smi} tool. The performance estimation from MPS then sets the corresponding speedup value to 0 before feeding the job information to \sol{} optimizer. For instance, if job $J_1$ cannot execute on \texttt{1g.5gb} slice, then the predictor sets $f_1(1) = 0$. The central controller maintains a ``maximum spare slice'' record for each GPU based on the memory constraints of its current jobs. It means that when re-partitioning the GPU, the maximum slice it can spare for a new job. When a job arrives in the queue with a memory limit, the controller will only consider GPUs whose ``maximum spare slice'' can satisfy the job memory constraint.\vspace{1.mm}

\vspace{2mm}\noindent\textbf{Quality-of-Service (QoS).} The user may specify a minimum slice size that the job can execute on so that the MIG slice provides enough performance to meet the QoS constraints. \sol{} deals with this constraint similar to the job memory constraint, the central controller will only send it to GPUs that can squeeze out a new slice satisfying QoS. 

\vspace{2.mm}
\noindent\textbf{Multi-instance jobs.} In special cases, one job may spawn multiple instances of the same workload to run in parallel, such jobs naturally fit on multi-instance GPUs. \sol{}'s performance predictor only runs for one instance of the job, then spawns all job instances on other GPUs using the profiled job information. The spawned instances do not need to be MPS profiled anymore -- \sol{} directly starts the optimizer.
\vspace{-1mm}

\subsection{Implementation}
\label{sec:implement}
\sol{}'s implementation is built upon MPS and MIG APIs that we develop. \revision{Each GPU runs in MIG mode all the time because switching MIG mode on and off incurs extra overhead. When the GPU needs to run in MPS mode, it changes its partition to \texttt{7g.40gb} and runs MPS on top of the \texttt{7g.40gb} MIG slice. This capability to run MPS on top of MIG is supported by NVIDIA~\cite{nvidia-mig}. During MPS, the GPU keeps an MPS control daemon in the background. To connect a job to the MPS daemon, we pass the \texttt{CUDA\_MPS\_PIPE\_DIRECTORY} variable to the job, which points to the same variable value specified by the daemon. To set the MPS level, we pass another variable \texttt{CUDA\_MPS\_ACTIVE\_THREAD\_PERCENTAGE} with the MPS level percentage as a value to the job.} The MIG API is more involved and richer as configuring the GPU from one MIG partition to another involves a series of commands to destroy compute and GPU instances, then create new GPU and compute instances. To ensure a job starts on the correct MIG slice, we also need to retrieve the UUID using \texttt{nvidia-smi} commands, as this UUID varies across different MIG devices and different GPUs. 
\revision{We use an automated script to collect the MIG device UUID for each partition and stored as lookup tables (only needed once). To assign a job to a particular MIG slice, we pass the \texttt{CUDA\_VISIBLE\_DEVICES} variable with the corresponding UUID to the job.} We have integrated these commands into Python function calls using the \texttt{subprocess} module. 


\revision{For each job submitted to the system, because all the MIG slice assignments and MPS control tasks are implemented by passing environment variables to the job, the user does not need to make additional code changes.} \sol{}'s server API (Fig.~\ref{fig:desi_1}) hosts a trained U-Net model in TensorFlow, and a partition optimizer utility. The GPU nodes do not communicate with each other, but they continue to update their status (i.e., job completion, current partition, MPS start/finish) to the central controller via TCP, so that the controller can decide the appropriate location for the next job.

\section{Methodology}
\label{sec:metho}

 
\noindent\textbf{Evaluation Setup.} We conduct real-system evaluations of \sol{} on an experimental testbed with four nodes, each node is equipped with 2 AMD EPYC 7542 CPUs and 2 NVIDIA A100-PCIe-40GB GPUs -- thus, 8 A100 GPUs in total. Note that one A100 GPU is reported to be comparable to 3 NVIDIA V100 GPUs or 10 NVIDIA P100 GPUs for datacenter applications~\cite{nvidia-a100}. With 56 MIG slices in total, our testbed can serve up to 56 jobs at any given time.

We also perform an extensive simulation-based evaluation to test \sol{}'s effectiveness on a 40-GPU A100 cluster. Simulation results are particularly of high significance since they show that \sol{}'s benefits are not limited to small-scale systems. In Sec.~\ref{sec:eval}, we conduct the simulation for 1000 different trials with a unique seed each time and report the results with violin plots and error bars.\vspace{1.5mm}

\noindent\textbf{Workloads.} \sol{}'s evaluation is driven by a job trace that emulates production behavior, in particular, our evaluation job trace is modeled after the most recently released and publicly available production GPU job trace for reproducibility and enhancement (Helios Trace~\cite{hu2021characterization}). For testbed experiments, we generate a 100-job mix that mimics the job execution time from the original trace when running on unpartitioned A100 GPUs. To accommodate the GPU hour time constraint, we limit the maximum job duration to be within 2 hours, which is approximately the 90$^{th}$ percentile execution time of the Helios Trace. Note that a 2-hour job on A100 GPUs could execute for 5 hours on smaller MIG slices, thus this limit helps us guarantee the completion of one set of experiments within a day. The job arrival follows a Poisson distribution with a $\lambda$ of 60 seconds. Poisson distribution is widely used to model job arrival in multiple previous works~\cite{narayanan2020heterogeneity,gu2019tiresias,ousterhout2017flexplane}. 

For simulator evaluations, we use the same trace to generate a 1000-job mix for each trial and use a $\lambda$ of 10 seconds for the Poisson distribution. The $\lambda$ parameter is also swept over a range of values to model different job arrival intensities in different situations. When repeating the simulation for 1000 trials, the job generation is fully randomized with different job mixes, arrival orders, and execution times.

\begin{table}
\small
\centering
\caption{Workloads used to evaluate \sol{}.}
\vspace{-0.2cm}
\scalebox{.9}{
\begin{tabular}{ccc}
\toprule
\textbf{Model} & \textbf{Batch Sizes} & \textbf{Application}\\ 
\midrule
\midrule
ResNet50~\cite{he2016deep} & \makecell{64, 128,\\256, 512} & \makecell{Image classification\\with residual learning}\\
\hline
MobileNet~\cite{howard2017mobilenets} & \makecell{64, 128,\\256, 512} & \makecell{Image classification\\on lightweight model}\\
\hline
BERT~\cite{devlin-etal-2019-bert} & 2, 4, 6, 8 & \makecell{Sentiment analysis of the\\IMDB movie reviews}\\
\hline
Transformer~\cite{vaswani2017attention} & \makecell{16, 32,\\64, 128} & \makecell{Time series prediction of\\engine noise measurement}\\
\hline
DeepSpeech~\cite{amodei2016deep} & \makecell{2, 4,\\8, 16} & \makecell{Automatic speech recognition\\of the LJSpeech dataset}\\
\hline
Embedding~\cite{pennington2014glove} & \makecell{64, 128,\\256, 512} & \makecell{Word embedding model for\\message topic classification}\\
\hline
Graph NN~\cite{gilmer2017neural} & \makecell{64, 128,\\256, 512} & \makecell{Property prediction of quantum\\chemistry molecular graphs}\\
\hline
CycleGAN~\cite{zhu2017unpaired} & \makecell{1, 2, 3, 4} & \makecell{Learning of mapping for\\image-to-image translation}\\

\bottomrule
\end{tabular}}
\label{table:metho}
\vspace{-0.6cm}
\end{table}

We use various types of deep learning (DL) workloads because the recent advancement in DL algorithms has made them popular in scientific research and production datacenters~\cite{shin2021revealing,wu2019performance,samsi2021supercloud}. We uniformly sample the DL model and training batch size from Table~\ref{table:metho}. These workloads come from Hugging Face~\cite{huggingface} and the \texttt{keras.io} repository~\cite{kerasio}. The application domains include computer vision, language modeling, speech recognition, and scientific computing, and have distinctive DL operators including CNN, RNN, and embedding tables. \textit{Disjoint sets of jobs mixes were used for training and testing.}\vspace{2mm}

\noindent\textbf{Competing Techniques.} As \sol{} is the first work to exploit MIG features for datacenter operation, we devise two intuitive competing techniques and one oracle scheme as below.

\textbf{\base{}.} This scheme does not perform MIG partition on A100 GPUs, reflecting the default GPU usage scenario in datacenters. It is simple to operate: when a system upgrades its GPU hardware to A100s, the system operator can manage them just like the previous GPU generation. 

\textbf{\static{}.} This scheme partitions all A100 GPUs into a fixed configuration that does not change over time. It is a straightforward way to manage MIG-enabled GPUs as a recent work Abacus~\cite{cui2021enable} has deployed static partitions of (\texttt{4g}, \texttt{2g}, \texttt{1g}) on their A100 GPUs. However, the best MIG configuration changes when running different job traces. To make sure we always use the optimal static partition when comparing against \sol{}, we exhaustively evaluate all possible MIG configurations offline and choose \textit{the best static partition}. Thus, the scheme is called optimal static (\static{}). 

\textbf{\orcl{}.} This is similar to \sol{} except that it uses oracle information about job profiles of MIG slice speedups, which are collected offline before execution. Hence, it does suffer from profiling overhead and prediction inaccuracies.\vspace{2mm}

\revision{Our evaluation results for \static{} and \orcl{} schemes do not include any profiling/switching overhead (ideal results), but our \sol{} results include its overhead for conservative performance improvement reporting. \static{} is the ``best static MIG configuration'' which works the best on average across all the job mixes (a single configuration). \orcl{} finds the best dynamic MIG configuration - different for different mixes. Therefore, \sol{} can be expected to outperform \static{} sometimes, but not \orcl{}.}\vspace{2mm}

\noindent\textbf{Metrics.} As discussed and defined earlier in Sec.~\ref{back:metrics}, we use three widely-used figures of merit: \textbf{average job completion time (JCT)}, \textbf{makespan} and the \textbf{system throughput (STP)}.  

\section{Evaluation}
\label{sec:eval}

\subsection{Real System Evaluation and Analysis}

First, we present results and analysis on a real cluster to demonstrate the effectiveness of \sol{} and derive insights.

\begin{figure}[t]
    \centering
    \includegraphics[scale=0.50]{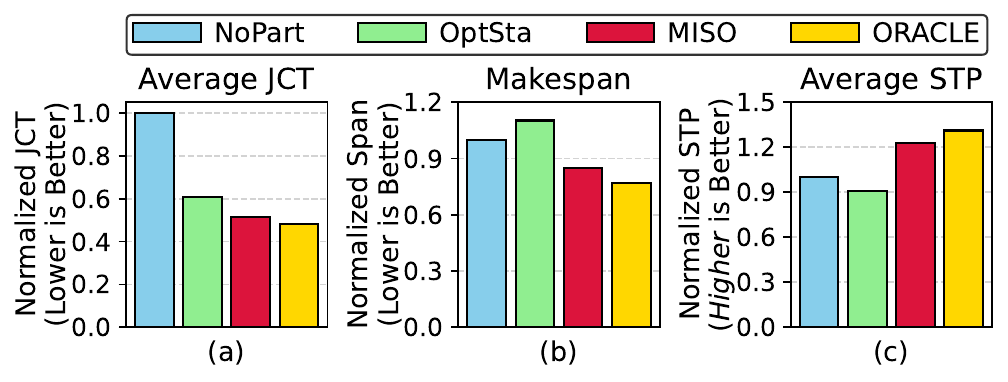}
    \hrule
    \vspace{-0.4cm}
    \caption{Performance comparison with competing techniques. All values are normalized to \base{}.}
    \vspace{-0.3cm}
    \label{fig:eval_exp_1}
\end{figure}

\vspace{1.5mm}
\noindent\textbf{Job completion time, makespan, and system throughput. } Fig.~\ref{fig:eval_exp_1} shows the average job completion time for different competing strategies, normalized to unpartitioned (\base{}) strategy. Recall that the unpartitioned strategy does not create MIG instances to co-locate jobs. We make several observations.

First, the optimal static partitioning (\static{}) outperforms unpartitioned scheme by 39\% (the absolute average JCT for \base{} is 40 minutes) (Fig.~\ref{fig:eval_exp_1}(a)). Recall that \static{} determines the optimal MIG instance partitioning via an exhaustive offline search process. The same partition is assigned to all the GPUs in the cluster, and this partitioning leads to better JCT for all jobs on average. Second, we observe that \sol{} significantly outperforms even this offline strategy by 16\%, even though \sol{} does not utilize any oracle information or require offline processing for decision-making. This is because \sol{} determines the GPU resource partition dynamically and specifically targets a given job mix. It also adjusts its partition as the job mix changes (e.g., arrival of new jobs, completion of existing jobs). Finally, our results show that \sol{} achieves similar performance to the Oracle strategy, which is dynamic and utilizes futuristic information. 

Fig.~\ref{fig:eval_exp_1}(b) and (c) reflect similar trends for the other two figures of merit: makespan and system throughput. \sol{} shortens the makespan by 23\% over optimal static partitioning and is within 10\% of the \orcl{} strategy. Similarly, \sol{} increases the system throughput by 35\% over \static{} and stays within 7\% of the \orcl{} strategy. \static{} outperforms the no-partition strategy in terms of JCT but performs worse than the no-partitioning strategy in terms of system throughput and makespan. This is because a few long-running jobs cannot access extra GPU resources when they are uncontested in \static{}, so they become straggler jobs with a long makespan. Nevertheless, \sol{} outperforms both \base{} and \static{} strategies in all aspects and is similar to the Oracle strategy. 

\begin{figure}[t]
    \centering
    \includegraphics[scale=0.50]{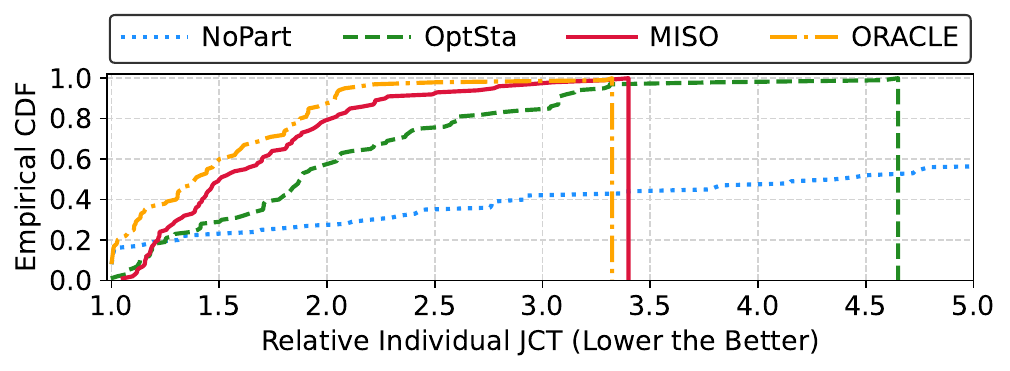}
    \hrule
    \vspace{-0.4cm}
    \caption{CDF of relative JCT of individual jobs. Each job's JCT is normalized to JCT when running on exclusive A100 GPU without queuing delay. The vertical line represents the maximum.}
    \vspace{-0.4cm}
    \label{fig:eval_exp_2}
\end{figure}

While Fig.~\ref{fig:eval_exp_1} confirms that \sol{} outperforms competing techniques, it only provides the average improvement across jobs. Next, we provide deeper quantitative evidence to demonstrate \sol{}'s effectiveness. Fig.~\ref{fig:eval_exp_2} shows the relative JCT for all jobs in the trace compared to their isolated, interference-free execution on the full GPU -- represented as the cumulative distribution function (CDF) of relative JCT for all competing techniques. This result confirms that overall average improvement in JCT (Fig.~\ref{fig:eval_exp_1}(a)) is not a result of \sol{}'s aggressive attention to certain jobs. In fact, Fig.~\ref{fig:eval_exp_2} shows that \sol{} consistently provides improvements for all jobs compared to all schemes. Similar to the Oracle strategy, 50\% of \sol{}'s jobs experience within 1.5$\times$ of the ideal JCT they can possibly have without sharing and queuing, while for \base{} and \static{}, this portion is less than 30\%. Next, we dig deeper to understand the reason behind \sol{}'s strength.

\begin{figure}[t]
    \centering
    \subfloat[][]{\includegraphics[scale=0.46]{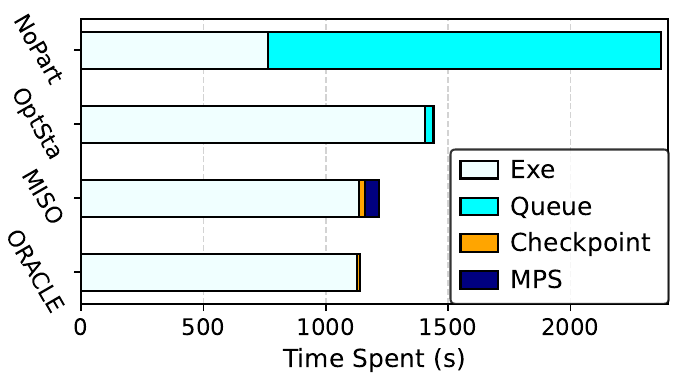}}\hfill
    \subfloat[][]{\includegraphics[scale=0.46]{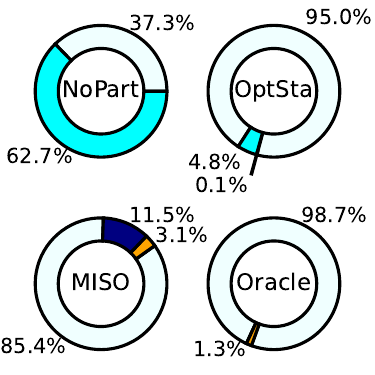}}\hfill    
    \hrule
    \vspace{-0.4cm}
    \caption{Breakdown of stages during the entire job life cycle. (a) shows the absolute time and (b) shows the percentage time.}
    \vspace{-0.4cm}
    \label{fig:eval_exp_3}
\end{figure}

\vspace{2mm}
\noindent \textbf{Why does \sol{} perform effectively? } We use two different experimental pieces of evidence to demonstrate the key sources of \sol{}'s effectiveness.  First, Fig.~\ref{fig:eval_exp_3} shows the breakdown of average job completion time spent in various stages for different competing schemes. As expected, jobs in the no-partition scheme spend over 60\% of their total time in the queue because of the unpartitioned GPU resources. While the jobs benefit from running colocation-free on these GPUs, ultimately, their queue wait time negatively affects the overall JCT. \static{} reduces the queue wait time by allowing effective co-locations, but because of its static nature, \static{} still remains sub-optimal and the jobs spend about 5\% of the time in the queue due to limited processing capability. Note that \static{} also migrates jobs from small slices to larger slices upon availability, but the checkpointing overhead is negligible (0.1\%). In contrast, \sol{} and \orcl{} completely eliminates queue wait time, providing evidence for their outstanding job processing power, and capability to support larger user bases. This is realized by incorporating dynamic partitioning across different GPUs depending upon the co-located job mix, instead of one single static partition across all GPUs in the cluster. However, \sol{} incurs extra checkpointing overhead because it requires jobs to run in MPS mode to estimate the optimal GPU partition. The job is still progressing towards completion during MPS mode, thus \sol{} is able to keep up with the pace of \orcl{} even though MPS accounts for 12\% of the time. 

\textit{This result also highlights the importance of \sol{}'s approach of using MPS mode to reduce MIG configuration explorations} -- this reduces the needed checkpoint to only 3\% in Fig.~\ref{fig:eval_exp_3}(b). If we do not start with MPS mode but choose to exhaustively profile the job speedups in MIG, this fraction grows to more than 20\% while jobs also experience significant idle periods during this process. This means that frequent checkpoints needed to explore different MIG partitions to determine a near-optimal partition is time prohibitive -- hence, highlighting the importance of MPS to MIG translation.

\begin{figure}[t]
    \centering
    \includegraphics[scale=0.51]{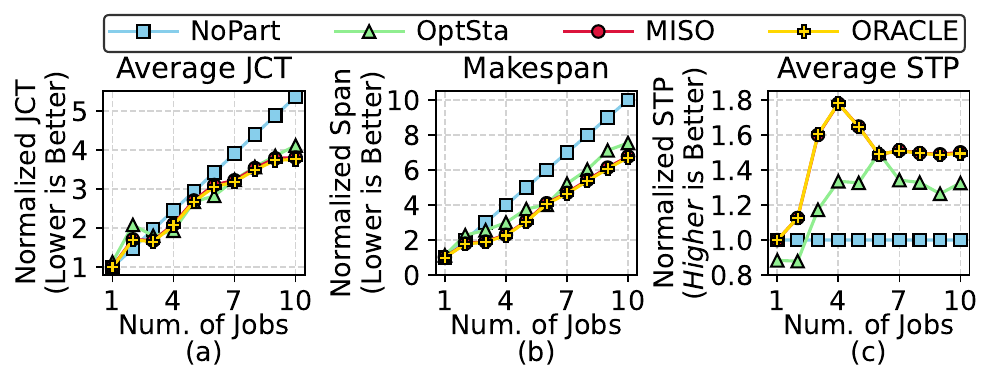}
    \hrule
    \vspace{-0.4cm}
    \caption{Scheduling more vs. less jobs on a GPU.}
    \vspace{-0.4cm}
    \label{fig:eval_exp_4}
\end{figure}

\vspace{2mm}Next, we conduct a single-GPU experiment to show \sol{}'s effectiveness as we increase the number of jobs scheduled for the GPU: we conduct 10 trials with increasing job number from 1 to 10, and each job lasts for 10 minutes on an exclusive A100 GPU. We show the results in Fig.~\ref{fig:eval_exp_4}, where all metrics are normalized to the 1-job \base{} trial. Because \base{} processes the jobs one by one, its average JCT and makespan follow a linear trend as the number of jobs increases, and its system throughput remains 1 due to no GPU sharing. First, we observe that the difference between \sol{} and \base{} broadens as the number of jobs increases, meaning \sol{} is more capable of processing heavier workloads. Second, the \static{} scheme could even occasionally outperform \sol{} in JCT, this shows that \static{} is a highly competitive scheme for some job mixes. However, in a system with a large number of GPUs and jobs, it is unlikely that every GPU can receive a job mix that matches well with the same static partition -- \sol{} resolves this issue with its job-mix-specific partition optimization at each GPU. Finally, almost all \sol{} and \orcl{} data points overlap in Fig.~\ref{fig:eval_exp_4}, meaning \sol{} has found the oracle partition for most job mixes. 

\begin{figure}[t]
    \centering
    \includegraphics[scale=0.50]{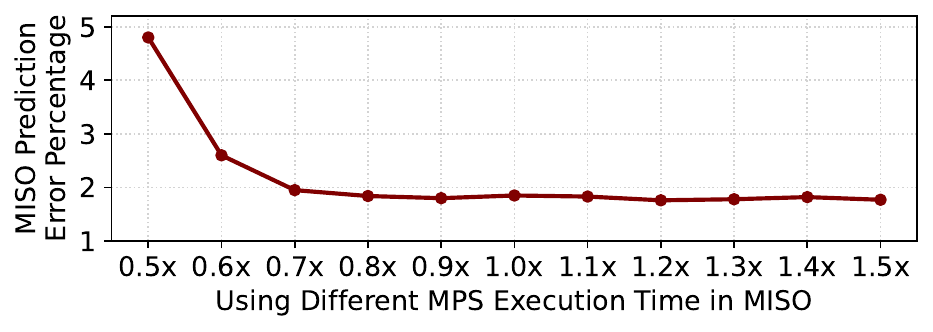}
    \hrule
    \vspace{-0.4cm}
    \caption{\revision{\sol{}'s prediction error when changing the MPS profiling time. 1$\times$ represents \sol{}'s current MPS time} \todo{at 10 seconds per MPS level (Sec.~\ref{sec:desi1})}}
    \vspace{-0.4cm}
    \label{fig:eval_exp_7}
\end{figure}

\vspace{2mm} 

\revision{\noindent \textbf{Benefit from longer MPS execution time yields diminishing returns, and \sol{} provides a significant advantage over the MPS-only approach.} Recall that \sol{} leverages brief MPS-mode execution to estimate the optimal MIG partition. Fig.~\ref{fig:eval_exp_7} shows the effects of increasing and decreasing \sol{}'s MPS profiling time. When the MPS time is cut to half (0.5$\times$), the prediction error becomes much higher. But further increasing MPS profiling time only yields diminishing returns in prediction accuracy. At 1.5$\times$ MPS profiling time, we have even observed a 4\% performance degradation in JCT, this is because the system does not have accuracy benefit from the longer MPS time, but experiences longer inefficient execution in MPS compared to running on optimal MIG partitions.} \todo{Later in Sec.~\ref{sec:eval_sim} we will also show that \sol{} is tolerant to a larger prediction error.}

\begin{figure}[t]
    \centering
    \includegraphics[scale=0.50]{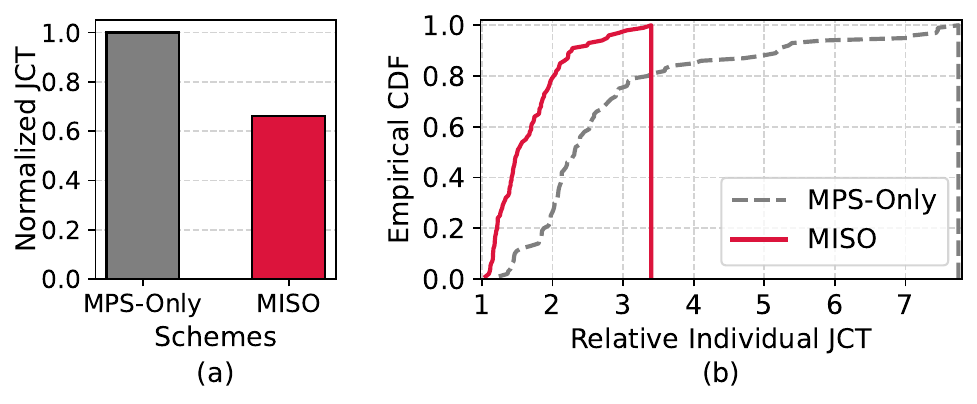}
    \hrule
    \vspace{-0.4cm}
    \caption{\revision{Comparing \sol{} against an MPS-only baseline. (a) shows the average JCT normalized to MPS-only. (b) shows the CDF of relative JCT of individual jobs compared to exclusive execution on full GPU.}}
    \vspace{-0.4cm}
    \label{fig:eval_exp_6}
\end{figure}

\revision{\sol{} achieves significant performance gains from the unpartitioned GPU baseline. In Fig.~\ref{fig:eval_exp_6}, we compare \sol{} against an MPS-only baseline partition to show that \sol{}'s benefits stem from intelligently partitioning the GPU resources using the MIG technology. The MPS-only scheme partitions each GPU's SM into three equally sized portions (limiting to three because more partitions lead to worse performance and out-of-memory error), and co-locates the jobs on these MPS partitions. Fig.~\ref{fig:eval_exp_6} (a) shows that \sol{} improves the average JCT by 35\% compared to the MPS-only baseline. Fig.~\ref{fig:eval_exp_6} (b) shows the relative JCT for individual jobs (same as Fig.~\ref{fig:eval_exp_2}) have shorter JCT when running on \sol{} -- 80\% of jobs have less than 2$\times$ JCT degradation compared to exclusive A100 execution on \sol{} while the corresponding portion is 30\% on MPS-only.}

\subsection{Simulation Evaluation and Analysis}
\label{sec:eval_sim}
Our simulation evaluation tests \sol{}'s effectiveness under different scenarios and at a larger scale (40 GPUs, 1000 jobs) that is cost-prohibitive on real systems. 

\begin{figure}[t]
    \centering
    \includegraphics[scale=0.50]{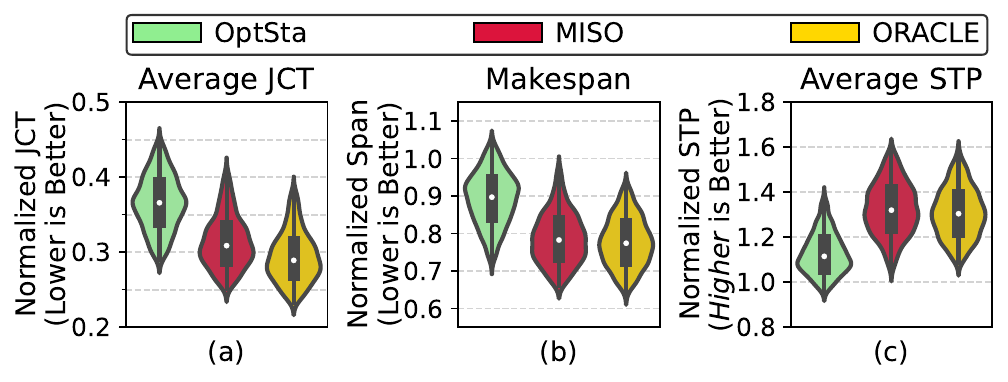}
    \hrule
    \vspace{-0.4cm}
    \caption{Violin plot of the results during the 1000 repetitions.}
    \vspace{-0.4cm}
    \label{fig:eval_sim_1}
\end{figure}

\vspace{2mm}
\noindent\textbf{Job completion time, makespan, and system throughput. } Our evaluation particularly focused on validating the consistent performance improvements by forcing each simulation run to start with different initial conditions (different job generation, arrivals, lengths). Hence the results yield different magnitudes of performance improvement. We use violin plots in Fig.~\ref{fig:eval_sim_1} to capture this difference. For each initial condition, we normalize the measurements of all techniques over \base{}. 

Our evaluation confirms that \sol{} indeed provides a significant improvement over competing schemes averaged over all runs, and stays close to \orcl{}'s improvement results -- not just the median/min/max, but throughout the full distribution. \sol{} provides about 70\%, 20\%, and 30\% median improvement over \base{} in terms of JCT, makespan, and system throughput. These improvements are amplified in the large-scale system compared to real system evaluation, showing \sol{}'s scalability. We have confirmed that when setting all simulation parameters to be the same as real system evaluation, they yield similar results.

\begin{figure}[t]
    \centering
    \includegraphics[scale=0.48]{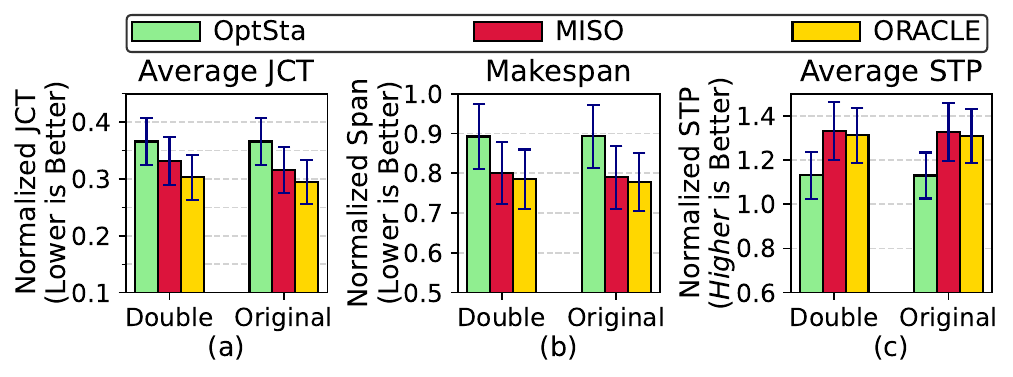}
    \hrule
    \vspace{-0.4cm}
    \caption{Sensitivity to checkpointing overhead.}
    \vspace{-0.4cm}
    \label{fig:eval_sim_2}
\end{figure}

\begin{figure}[t]
    \centering
    \includegraphics[scale=0.48]{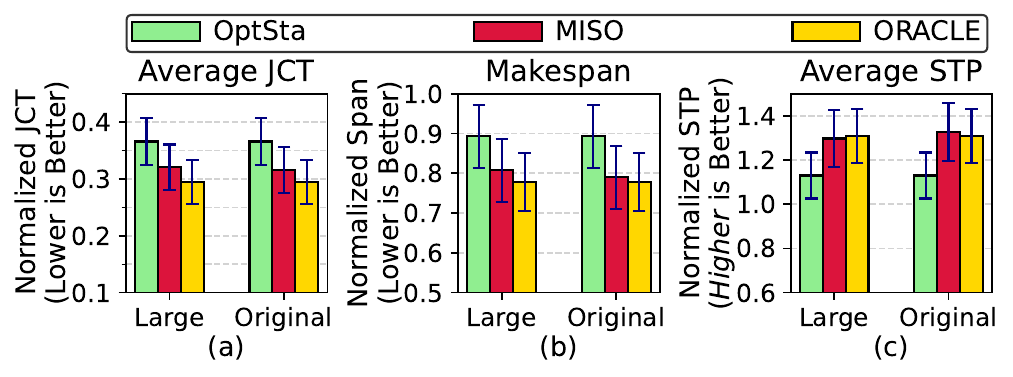}
    \hrule
    \vspace{-0.4cm}
    \caption{Sensitivity to performance prediction error.}
    \vspace{-0.4cm}
    \label{fig:eval_sim_3}
\end{figure}

\vspace{2mm}
\noindent\textbf{Sensitivity to checkpointing overhead, performance model prediction error, and inter-arrival rate. } Finally, we evaluate the sensitivity of different system and design parameters on \sol{}'s effectiveness. Recall that \sol{} operation relies on (1) checkpointing, and (2) performance prediction from MPS to MIG mode. \sol{} incurs checkpointing overhead during MPS profiling and MIG re-partitioning. Our results (Fig.~\ref{fig:eval_sim_2}) confirm that this overhead does not impact \sol{}'s benefits even when the checkpointing overhead doubles -- presumably with a hypothetical system of much slower memory bandwidth or jobs that are much larger in size. Fig.~\ref{fig:eval_sim_3} shows that even when the performance prediction model is just trained for a couple of epochs with a large prediction error (error from 1.7\% to 9\%), \sol{} still provides a comparable improvement over non-partitioned GPUs without a fine-tuned model. 

\begin{figure}[t]
    \centering
    \includegraphics[scale=0.48]{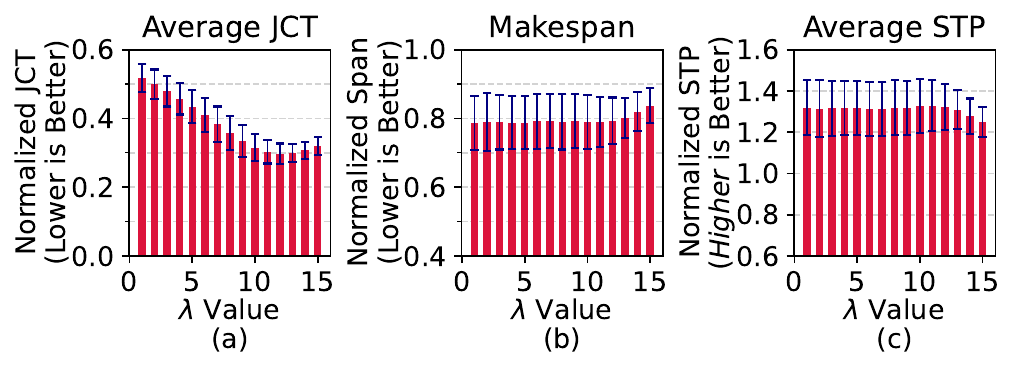}
    \hrule
    \vspace{-0.4cm}
    \caption{Sensitivity to arrival rate ($\lambda$ unit: seconds).}
    \vspace{-0.4cm}
    \label{fig:eval_sim_4}
\end{figure}

Finally, we show that \sol{} remains effective as the job inter-arrival time changes (Fig.~\ref{fig:eval_sim_4}). This test was performed to simulate systems with different loads. A low inter-arrival time (small $\lambda$) requires \sol{} to profile and adjust the MIG partitions more frequently and oversubscribes the GPUs. Therefore, its relative JCT performance degrades. In spite of that, \sol{} still maintains its improvements in makespan and system throughput. \sol{} continues to provide 30\% to 50\% of average JCT improvement, more than 15\% of makespan improvement, and more than 25\% higher system throughput over \base{} across a wide range of inter-arrival rates.\vspace{-0mm}

\section{Related Work}
\label{sec:relate}



Prior extensive works on co-locating workloads on CPU-based servers~\cite{delimitrou2013paragon,zhang2014smite,blagodurov2015multi,zhu2016dirigent,kasture2015rubik,wang2017swap,kasture2014ubik,el2018kpart,xiang2018dcaps,xu2018dcat,park2018hypart,park2019copart,margaritov2019stretch,patel2020clite,yi2022mt2} do not provide a solution to GPU-specific co-location challenges (e.g., different architecture organization, resource sharing granularity, and allowable resource partitions). Consequently, multiple works have investigated GPU-specific sharing. 

In particular, Clockwork~\cite{gujarati2020serving}, S$^3$DNN~\cite{zhou2018s}, and DART~\cite{xiang2019pipelined} design CUDA stream schedulers for multiple DNNs to share a GPU at the operator level. TimeWall~\cite{amert2021timewall}, Gandiva~\cite{xiao2018gandiva}, Gandiva-fair~\cite{chaudhary2020balancing}, and Antman~\cite{xiao2020antman} use GPU time-sharing to improve resource utilization during idle job cycles. Space sharing is suitable when a single application cannot efficiently use the entire GPU, which is addressed by Gavel~\cite{narayanan2020heterogeneity} and Gslice~\cite{dhakal2020gslice} in MPS sharing mode. Many other works have addressed various areas in GPU sharing including communication, memory allocation, and latency sensitivity~\cite{lee2022gshare,ranganath2021mapa, 10.1145/3503222.3507721,10.1145/3508036,10.1145/3503221.3508423}. However, none of the above works addresses the challenges and limitations of using MIG-enabled GPU sharing which, as we discussed in Sec.~\ref{sec:bkgd} and~\ref{sec:motiv}, has its own specific challenges and benefits. Recently, Abacus~\cite{cui2021enable} and Zahaf et al.~\cite{zahaf2021contention} have used MIG-enabled GPU for their experimental evaluation, but they rely on a naive static resource partitioning -- understandably so since their focus is not to improve performance via determining the best partition of resources.  \revision{GPU sharing has also been studied on the device memory, including concurrent query processing~\cite{wang2014concurrent} and virtual memory management for co-located applications~\cite{wang2014gdm}. These works have pushed forward the research field till the recent MIG technology appears. \sol{} is built upon MIG's hardware-supported memory sharing and isolation.} In summary, \sol{} is the first work to address various challenges in operating a MIG-enabled GPU datacenter and provide solutions for improving job completion time and system throughput.
\section{Discussion}
\label{sec:discuss}

\noindent\textbf{{GPU resource partitioning beyond NVIDIA A100 GPUs.}} {We anticipate that hardware and software support for GPU resource partitioning will become prevalent as single GPU nodes become more powerful and all the architectural resources within a single GPU cannot be maximally utilized by a single application all the time. NVIDIA's Ampere architecture (A100) is the first commercially available GPU to provide this capability via MIG technology. NVIDIA's next-generation architecture, Hopper, will continue to offer MIG support~\cite{nvidia-h100}. Other GPU vendors also realize this opportunity and are working toward providing similar support. For example, AMD's Compute Unit (CU) masking library in the ROCm (Radeon Open Compute) stack, will potentially allow partitioning of the CUs similar to NVIDIA's MPS~\cite{otterness2021exploring} and similar approach is anticipated from Intel~\cite{intel}.}




\vspace{2mm}
\noindent\textbf{{Scalability w.r.t. the number of partition combinations in future MIG-based GPUs.}} {As GPUs evolve, it is possible that future generation GPUs may have more MIG slices, and hence, more number of combinations than today (currently, 18 combinations). There are two major implications: (1) a larger number of MIG slice types could affect \sol{}'s performance prediction accuracy for different MIG slices, and (2) \sol{}'s partition optimizer algorithm needs to account for a larger number of partition combinations. Fortunately, \sol{}'s design is reasonably robust to these issues. For the first issue, our sensitivity study (Fig.~\ref{fig:eval_sim_3}) shows that \sol{} can tolerate some prediction errors in its model (from 1.7\% to 9\%) and still provide significant improvements. Furthermore, we can leverage transfer learning to improve the accuracy of our models as the number of combinations increases. For the second issue, we experimentally measured that Algorithm~\ref{algo:miso} finishes within \SI{80}{\milli\second} even with 10$\times$ the number of combinations (total of 180 combinations). This is because the partition optimizer runtime scales linearly with the number of combinations for a given degree of co-location. Even with a 100$\times$ increase, the optimizer finishes within a second, and its latency is overlapped with the execution of workloads.}



\vspace{2mm}
\noindent\textbf{{Future work and opportunities enabled by \sol{}.}} {\sol{} demonstrates effective co-location of multiple jobs for higher throughput. Each single GPU node can be treated as a combination of different small GPUs (i.e., multiple heterogeneous partitions within a GPU). The cloud computing providers and HPC cluster managers may expose different partitions of a large GPU directly to users as a job allocation unit. \sol{} will enable cloud computing users to leverage \sol{}'s performance predictor to estimate a job's performance on different sub-GPUs, and request those partitions accordingly. \textit{Finally, we hope that \sol{} can also help cloud compute providers appropiately price their sub-GPUs (in terms of monetary cost or core hours) as a single resource consumption unit and expose them as compute units for rent.}}


\section{Conclusion}
\label{sec:conclude}

In this paper, we presented \sol{}, a technique to leverage the MIG functionality on NVIDIA A100 GPUs to dynamically partition GPU resources among co-located jobs. \todo{\sol{} deploys a learning-based method to quickly find the optimal MIG partition for a given job mix running in MPS.} \sol{} \todo{is evaluated using a variety of deep learning workloads and} achieves an average job completion time that is lower than the unpartitioned GPU scheme by 49\% and is within 10\% of the Oracle technique.

\section*{APPENDIX}
\label{sec:appendix}
\begin{figure}[H]
    \centering
    \includegraphics[scale=0.44]{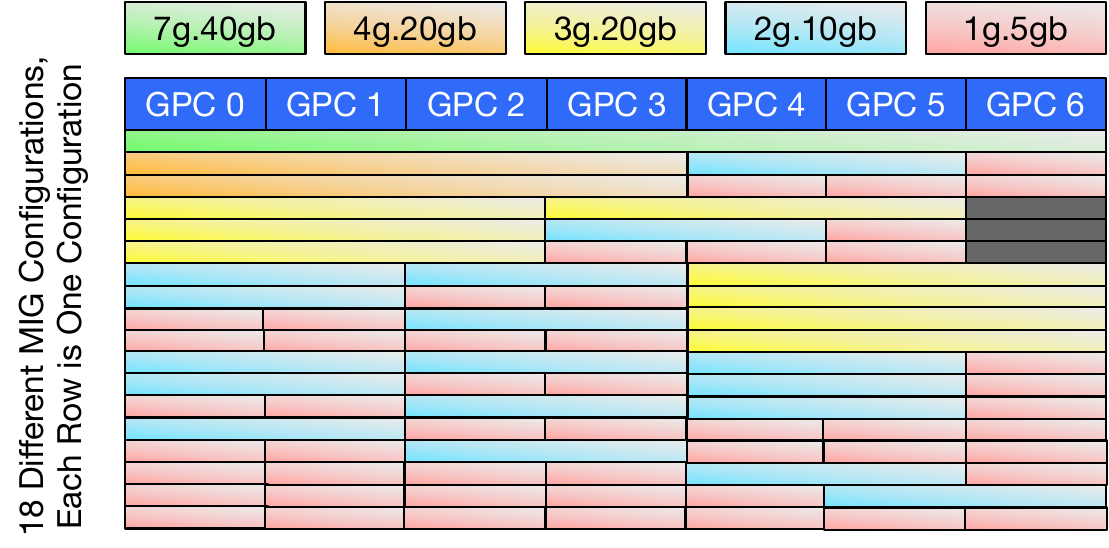}
    \vspace{2mm}
    \hrule
    \vspace{-0.4cm}
    \caption{\revision{All possible MIG configurations in A100, created according to NVIDIA's MIG user guide~\cite{nvidia-mig}.}}
    \vspace{-0.4cm}
    \label{fig:appendix}
\end{figure}

\setcounter{section}{1}
\revision{Fig.~\ref{fig:appendix} visually shows all 18 possible MIG configurations in an A100 GPU. Each row represents one configuration (e.g., the second row represents (\texttt{4g.20gb}, \texttt{2g.10gb}, \texttt{1g.5gb}.)}

\begin{acks}
\todo{We would like to thank anonymous reviewers and our shepherd (Purushottam Kulkarni) for their feedback which helped us improve this paper. This research was supported by NSF Award 1910601, NSF Award 2124897, and Northeastern University.} This research was also sponsored by the United States Air Force Research Laboratory and the United States Air Force Artificial Intelligence Accelerator and was accomplished under Cooperative Agreement Number FA8750-19-2-1000. The views and conclusions contained in this document are those of the authors and should not be interpreted as representing the official policies, either expressed or implied, of the United States Air Force or the U.S. Government. The U.S. Government is authorized to reproduce and distribute reprints for Government purposes notwithstanding any copyright notation herein.
\end{acks}

\bibliographystyle{unsrtnat}
\bibliography{reference}

\end{document}